\documentclass[sigconf,authorversion]{acmart}

\hypersetup{bookmarksdepth=-2}

\usepackage{algorithm2e}
\usepackage{booktabs}
\usepackage{listings}
\usepackage{lstlinebgrd}
\usepackage{multirow}

\definecolor{lightyellow}{RGB}{255,241,217}
\begin{document}
\title{Augmenting Source Code Lines with Sample Variable Values}

\author{Mat\'u\v{s} Sul\'ir}
\orcid{0000-0003-2221-9225}
\affiliation{%
  \institution{Technical University of Ko\v{s}ice}
  \streetaddress{Letn\'a 9}
  \city{Ko\v{s}ice}
  \postcode{042 00}
  \country{Slovakia}
}
\email{matus.sulir@tuke.sk}

\author{Jaroslav Porub\"an}
\affiliation{%
  \institution{Technical University of Ko\v{s}ice}
  \streetaddress{Letn\'a 9}
  \city{Ko\v{s}ice}
  \postcode{042 00}
  \country{Slovakia}
}
\email{jaroslav.poruban@tuke.sk}

\begin{abstract}
Source code is inherently abstract, which makes it difficult to understand. Activities such as debugging can reveal concrete runtime details, including the values of variables. However, they require that a developer explicitly requests these data for a specific execution moment. We present a simple approach, RuntimeSamp, which collects sample variable values during normal executions of a program by a programmer. These values are then displayed in an ambient way at the end of each line in the source code editor. We discuss questions which should be answered for this approach to be usable in practice, such as how to efficiently record the values and when to display them. We provide partial answers to these questions and suggest future research directions.
\end{abstract}

%
%

\ccsdesc[500]{Software and its engineering~Integrated and visual development environments}
\ccsdesc[500]{Software and its engineering~Software maintenance tools}

\copyrightyear{2018}
\acmYear{2018}
\setcopyright{acmlicensed}
\acmConference[ICPC '18]{ICPC '18: 26th IEEE/ACM International Conference on Program Comprehension }{May 27--28, 2018}{Gothenburg, Sweden}
\acmBooktitle{ICPC '18: 26th IEEE/ACM International Conference on Program Comprehension, May 27--28, 2018, Gothenburg, Sweden}
\acmPrice{15.00}
\acmDOI{10.1145/3196321.3196364}
\acmISBN{978-1-4503-5714-2/18/05}

\keywords{integrated development environment (IDE), source code augmentation, examples, variables, dynamic analysis}

\maketitle

\section{Introduction}

Source code of a computer program is inherently abstract since it contains a specification of all possible executions. Trying to understand a program only by reading its source code is often difficult. Consider the following excerpt from the method createNumber in class NumberUtils of Apache Commons Lang\footnote{\url{https://commons.apache.org/proper/commons-lang/}}:
\begin{lstlisting}
exp = str.substring(expPos + 1, str.length() - 1);
\end{lstlisting}

While the variable names give the programmer a hint about their meaning, it requires non-negligible mental effort to construct a mental model of the presented line. Now, consider the same line augmented with concrete values of variables:
\begin{lstlisting}
exp = str.substring(expPos + 1, str.length() - 1);
$\color{gray}\textbf{str:} "1.1E-700F" \ \textbf{expPos:} 3 \ \textbf{exp:} "-700"$
\end{lstlisting}

The programmer can now intuitively understand that the line extracts the exponent from a string containing a decimal number in the scientific notation. Furthermore, we can see that the removal of the last character is necessary because it contains the suffix ``F'' (float).

To perform an investigation of dynamic program properties, such as concrete values of variables, developers often use debuggers \cite{LaToza10developers}. For instance, newer versions of IntelliJ IDEA already display an augmentation similar to the one presented above. However, the use of a debugger requires additional effort from the programmer. First, we must manually choose appropriate breakpoint locations (although an approach by Steinert et al. \cite{Steinert09debugging} can simplify this). Next, the program must be executed and the debugger must be guided using the stepping operations to progressively reveal variable values. At one moment, only one state of the program is visible and the programmer cannot get an overview of multiple states. Finally, after these laborious tasks are performed, the obtained dynamic information disappears as soon as the given lines become out of scope or the debugging session is closed.

To tackle this problem, tracing tools and time-traveling debuggers record the program execution and then allow the programmer to explore any part of the program in any recorded execution state. Nevertheless, these tools still require the user to manually navigate a large trace to select the relevant part. The runtime information and the static source code view are separated, requiring the user to switch between two disconnected views.

There exist several approaches trying to integrate runtime information directly into the source code editor, but the kinds of information they display is often limited. Senseo \cite{Rothlisberger12exploiting} displays information such as lists of callers, callees and dynamic argument types. Tral\-fa\-ma\-do\-re \cite{Lefebvre12execution} is limited to argument and return values, IDE sparklines \cite{Beck13visual} to numeric variable types.

Other approaches -- Tangible code \cite{Mellis06tangible} and Pathfinder \cite{Perscheid10immediacy} -- display variable values inline, but they are trace-focused: the programmer browses a trace and manually selects a point in the history which should be displayed.

Although DynamiDoc \cite{Sulir17generating} collects sample values at runtime, the results are only generated documentation sentences, not an interactive approach. Kr\"amer et al. \cite{Kraemer14how} integrate variable values with source code in a live programming environment. Such environments suffer from performance and scalability problems though.

In this article, we introduce an approach to present sample values of variables to a developer in an ambient and unobtrusive manner. Executions of tests or the application itself by a programmer are partially recorded. Then, at the end of each line, a noneditable comment-like label is displayed in the IDE. Each such label shows the values of the local and instance variables read or written on the given line, during one sample execution of this line.

\begin{figure}
\includegraphics[width=\columnwidth]{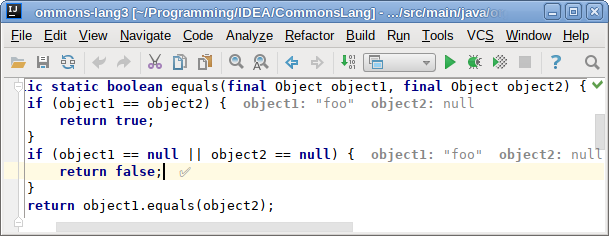}
\caption{Sample values augmentation in RuntimeSamp} \label{f:screenshot}
\end{figure}

A tool embodying this idea, RuntimeSamp, is implemented as a plugin for IntelliJ IDEA combined with a load-time Java bytecode instrumentation agent. For an illustration, see Figure~\ref{f:screenshot}. The source code of RuntimeSamp is available online\footnote{\url{https://github.com/sulir/runtimesamp}}.

While the approach itself seems simple, a number of interesting questions arose during the process of designing it. In this paper, we devise solutions to the encountered problems -- some naive and some more elaborate -- and try to foster future research in this area.

\section{Augmentation Approach}

In this section, we will present RuntimeSamp in more detail and discuss the individual questions.

\subsection{Object Representation}

Since the variable values are to be displayed in the text editor, they should be ideally represented on one line, using only a small amount of space. For primitive values, such as integers or floats, this is simple -- we use their traditional mathematical notation. The situation is more complicated for objects composed of many properties. Therefore, we formulate our first question: \textbf{Q1:} How to present complicated objects succinctly on a small space?

If we consider only purely textual representations, the most straightforward solution is to call the ``to string'' method (e.g., toString in Java) on a particular object. This is also the solution we used in RuntimeSamp (with a few exceptions, such as manually generating the array representation or shortening too long descriptions). This approach has its disadvantages, though. In many languages, generation of the string representation must be manually programmed for each class. Because it is useful almost exclusively for debugging purposes, developers often omit it. This leaves us with a default implementation similar to ``\texttt{MyClass@4a54c0de}'', which is obviously not useful.

It is trivial to recursively traverse all member variables of the given object down to primitive values or to the given depth and automatically produce a string in the form ``\{member$_1$: \{sub-member$_1$: value$_1$, sub-member$_2$: value$_2$, ...\}, ...\}.'' Nevertheless, such a string can be too long for the majority of objects. An interesting future research idea is to filter the list of member variables of each object to include only that relevant for comprehension. For example, we can exclude members representing purely implementation details; or include only frequently read or recently changed fields. Finding what constitutes a given member relevant and useful for comprehension remains an open question.

Regarding graphical and semi-graphical representations, a universal way would be to display a fully collapsed tree of all properties, represented by a clickable plus-sign ($\boxplus$) directly in the editor. The developer could expand it on demand and show the values of interest. On the other hand, this approach could affect the ambient nature of RuntimeSamp since it would require manual actions from the developer.

While there exist methods of graphical representation generation (e.g., the Moldable Inspector \cite{Chis15moldable}), they suffer from similar shortcomings as the classical toString() -- the implementation is manual, optional and not universal.

\subsection{Recording Moment Selection}

One variable can be displayed multiple times on each line, every time holding a different value. Thus the following question arises: \textbf{Q2:} When exactly should we capture the values of variables?

The following options are available:
\begin{itemize}
\item after reading/writing a variable for the first time on this line,
\item after the last reading/writing of it,
\item at the end of the line.
\end{itemize}
Note the value will be displayed at the end of the line. To match its visual position with its meaning, we selected the third option.

Our approach works with physical lines, so even if multiple statements are present on the same line, each variable is displayed only once in the augmentation.

An alternative would be to display a value directly next to each variable in the code (as in Tangible Code \cite{Mellis06tangible}). While this would improve spatial immediacy by lowering the distance between the variable occurrence and its value, the source code editor could easily become cluttered, especially in the case of more verbose object representations.

\subsection{Iteration Selection} \label{s:iteration-selection}

The most important problem regarding our approach stems from the fact that program execution is not sequential. Program constructs such as loops cause one source code line to be executed multiple times, each time in a possibly different context. Therefore, displaying an arbitrary sample value can cause confusion. Consider the following augmented source code excerpt, separately counting the sums of even and odd numbers:

\begin{lstlisting}[numbers=left,xleftmargin=1.7em]
for (int n : nums)  $\color{gray}\textbf{nums:} \{1, 2\} \ \textbf{n:} 1$
  if (n $\%$ 2 == 0)  $\color{gray}\textbf{n:} 1$
    even += n;  $\color{gray}\textbf{n:} 2 \ \textbf{even:} 2$
  else
    odd += n;  $\color{gray}\textbf{n:} 1 \ \textbf{odd:} 1$
\end{lstlisting}

Here, each line is augmented with the values recorded during the first time the particular line was executed. Up to the second line, everything seems to be consistent. However, at line 3, the value of \texttt{n} suddenly becomes 2, even though on the previous line it was 1. Clearly, the sample values at lines 1, 2 and 5 were recorded during the first loop iteration, while the augmentation at line 3 comes from the second one. This happened because the third line was not executed at all in the first iteration.

It is obvious that we can display only one iteration at a time. This leaves us with a question: \textbf{Q3:} How to decide which iteration to display?

The most elementary solution would be to insert a combo box at each line containing a looping construct (in our case, \texttt{for} at line~1) to enable a selection of a particular loop iteration from a list. A similar solution was already used, e.g., in compacted sequence diagrams \cite{Myers10utilizing}. However, this would require a nontrivial manual action from the programmer. Imagine selecting a relevant iteration from a list containing hundreds of items, without knowing any information about them besides the iteration numbers.

First of all, we need to provide a convenient way to select an iteration in which a particular line of interest was executed (an iteration which covers it). In RuntimeSamp, we set the ``line of interest'' to the line where the text cursor (caret) is currently located. For example, if the caret is on line 3, we show the second iteration:

\begin{lstlisting}[numbers=left,xleftmargin=1.7em,linebackgroundcolor={\ifnum\value{lstnumber}=3\color{lightyellow}\fi}]
for (int n : nums)  $\color{gray}\textbf{nums:} \{1, 2\} \ \textbf{n:} 2$
  if (n $\%$ 2 == 0)  $\color{gray}\textbf{n:} 2$
    even += n;$\rule[-0.35em]{0.13em}{1.2em} \ \color{gray}\textbf{n:} 2 \ \textbf{even:} 2$
  else
    odd += n;
\end{lstlisting}

On the other hand, if the cursor is on line 5, the first iteration is displayed:

\begin{lstlisting}[numbers=left,xleftmargin=1.7em,linebackgroundcolor={\ifnum\value{lstnumber}=5\color{lightyellow}\fi}]
for (int n : nums)  $\color{gray}\textbf{nums:} \{1, 2\} \ \textbf{n:} 1$
  if (n $\%$ 2 == 0)  $\color{gray}\textbf{n:} 1$
    even += n;
  else
    odd += n;$\rule[-0.35em]{0.13em}{1.2em} \ \color{gray}\textbf{n:} 1 \ \textbf{odd:} 1$
\end{lstlisting}

This way, we utilize the text cursor as an implicit pointer to the programmer's current focus point.

Of course, many lines can be covered by more than one iteration. The choice of the most relevant one is most probably task-dependent and remains an open research question.

Note that in Figure~\ref{f:screenshot}, the line with the \texttt{return false;} statement is augmented with a check mark (\checkmark). This is used for lines which were executed during the given iteration but do not contain any variables.

\subsection{Iteration Detection}

Although in the previous text, we intuitively worked with the notion of an ``iteration'', the situation is not always that simple. For instance, a loop can contain a call to a method defined in another file. An iteration statement is thus spatially disconnected from the place of the manifestation of its effect. Furthermore, such a loop can be located in a library or a system file which is not instrumented and recorded. Finally, many languages contain a variety of looping constructs which are difficult to recognize in the compiled bytecode. This leaves us with question \textbf{Q4:} How to define an iteration in a way which is easy to detect and present?

In RuntimeSamp, we used the notion of forward executions. As long as the program progresses forward without jumping back, we consider this one iteration, which we call a ``pass''. Since in RuntimeSamp, the recording and augmentation are line-oriented, we consider only jumps to another line, not the same one. Execution of another method starts a new pass; however, as soon as this method returns control to the caller, the original pass continues. Our approach is simple to implement using a local variable called \texttt{passId}, inserted into each method by instrumentation. This variable is assigned a new value on every method start and backward jump -- from a global variable \texttt{passCounter}, incremented on each read.

For an illustration, see the following code:
\begin{lstlisting}[numbers=left,xleftmargin=1.7em]
void caller() {
  int i = 1;
  callee();
  i = 2;
}
void callee() { doSomething(); }
\end{lstlisting}
Lines 2, 3, and 4 are a part of the pass with ID 1; line 6 pertains to pass ID 2.

\subsection{Collection Efficiency}

While the collection of all variable values in the whole program during the whole execution gives precise results, it is clearly not practical because of high time overhead and storage requirements. The following question arises: \textbf{Q5:} How to collect enough data for sample values presentation while keeping the overhead reasonable?

In RuntimeSamp, we aim to present sample values at the end of each line. Therefore, it is convenient to collect values at least once for every line being executed. In the current implementation, we use a simple logic. In a global array called \texttt{hits}, we keep the number of times each particular line of the program (identified by its \texttt{lineId}) was executed. As soon as a hard-coded ``hits per line'' limit is reached, we stop data collection for the given line.

\begin{algorithm}[b]
\SetKwFunction{SaveToDb}{SaveToDb}
On each method start do:\\
$passId \leftarrow 0$\;
\vspace{1ex}
Each time when the executing line is about to change from $currentLine$ to $nextLine$ do:\\
\If{hits[lineId] < HITS\_PER\_LINE}{
  $hits[lineId]$++\;
  \If{passId = 0}{
   $passId \leftarrow passCounter$++\;
  }
  \SaveToDb{fileName, currentLine, passId}\;
  \For{each variable read/written on currentLine}{
  	\SaveToDb{passId, name, value.toString()}
  }
}
\If{nextLine < currentLine}{
  $passId \leftarrow 0$\;
}
\caption{\textbf{Variable-recording instrumentation}} \label{a:instrumentation}
\end{algorithm}

When combined with the iteration detection approach described in the previous section, the result is depicted in Algorithm~\ref{a:instrumentation}. Note that to prevent the likely overflow of \texttt{passCounter}, we first reset the \texttt{passId} to 0 and obtain a new value from \texttt{passCounter} only when the data will be actually collected.

Clearly, the presented algorithm, which records the first n executions of each line, is not ideal. Consider the source code from section~\ref{s:iteration-selection} was executed with \texttt{nums} containing a thousand even items and one odd item as the last element. We would need to record the first iteration, then skip recording of 999 iterations and record the last one. In case the array was dynamically generated, a nontrivial prediction using dynamic analysis would need to be applied. Otherwise, we need to either capture all executions and sacrifice performance, or end up with incompletely recorded passes. In the example from section~\ref{s:iteration-selection}, with $HITS\_PER\_LINE$ in our algorithm set to 1, two passes are recorded: the first pass containing lines 1, 2, 3; and the second one containing only line 5.

We performed a preliminary performance evaluation of our approach. In Table~\ref{t:overhead}, we present time overhead of the instrumented runs when compared to plain, non-instrumented ones (zero means no overhead). The benchmarks come from DaCapo \cite{Blackburn06dacapo}, version 9.12-MR1. Auxiliary libraries and system classes were not instrumented. Beware our implementation is naive and can clearly be improved.

\begin{table}
\caption{Time overhead of instrumented runs} \label{t:overhead}
\begin{tabular}{cllll}
\toprule
\multirow{2}{*}{Hits per line} & \multicolumn{4}{c}{Overhead for benchmark} \\
~ & avrora & fop & h2 & xalan \\ \midrule
1 & 0.78 & 2.13 & 1.84 & 0.88 \\
2 & 0.89 & 2.37 & 2.07 & 0.93 \\
3 & 0.98 & 2.63 & 2.23 & 1.19 \\
\bottomrule
\end{tabular}
\end{table}

We may also consider collecting data only for a limited part of the program or an execution period. For example, the values of relevant variables could augment the source code when the program crashed due to an uncaught exception.

\subsection{Filtering of Variables}

A lot of data can visually clutter the editor and cause information overload. Thus our next question is \textbf{Q6:} Which variable values should be displayed and which not? Currently, we show all variables except same-named variables on the same line -- e.g., in the expression \texttt{this.var = var}, we show only one \texttt{var}. In the future, it is possible to devise other filtering patterns to achieve the optimal ratio between readability and completeness.

\subsection{Data Invalidation}

Our final question is \textbf{Q7:} When and how to invalidate the collected variable values? Currently, as a preliminary form of invalidation, we delete all data on any document edit. Of course, a more reasonable solution is to delete only data pertinent to statements dependent on the made changes. However, we can go even further: The augmentation of values which could be recomputed in a reasonable time (and without requiring additional input) should be updated in the editor, which would enable partial liveness.

\section{Conclusion}

We regard the source code view as the primary view of the programmer's interest. Therefore, we aim to deliver a ``feeling of runtime'' to the static source code by the means of augmenting source code editor lines with examples of concrete variable values. Our final goal is to make runtime information in the editor:
\begin{itemize}
\item ambient -- using implicit interaction information instead of requiring manual actions,
\item unobtrusive -- not requiring a change of the developer's tools or workflow, 
\item relevant -- thus aiming to leverage program comprehension.
\end{itemize}

Throughout the paper, we discussed several questions related to these properties. Still, many questions remain to be answered in future experiments. Particularly, an evaluation with human participants (industrial developers, students) should be performed, both to quantitatively confirm the utility of RuntimeSamp and to provide qualitative insights about the decisions made when designing it.

\begin{acks}
This work was supported by project KEGA 047TUKE-4/2016 Integrating software processes into the teaching of programming.
\end{acks}

\bibliographystyle{ACM-Reference-Format}
\bibliography{icpc}

\end{document}